\documentclass[12pt,preprint]{aastex}

\newcommand{\etal}{et~al.}
\newcommand{\fig}{Fig.~}
\newcommand{\mic}{\,$\mu$m}
\newcommand{\av}{A_V}
\newcommand{\water}{H$_2$O}
\newcommand{\cod}{CO$_2$}
\newcommand{\df}{F_{*}}
\newcommand{\sx}{_{\rm X}}
\newcommand{\nhmean}{\langle n_{\rm H} \rangle}

\slugcomment{Version date: \today}

\begin{document}

\title{OXYGEN DEPLETION IN THE INTERSTELLAR MEDIUM: \\ 
		IMPLICATIONS FOR GRAIN MODELS AND THE \\
		DISTRIBUTION OF ELEMENTAL OXYGEN\\ ~}
		
\author{D.~C.~B. Whittet}

\affil{New York Center for Astrobiology, and Department of Physics, Applied Physics \& Astronomy,
	Rensselaer Polytechnic Institute, 110 Eighth Street, Troy, NY 12180.}

\begin{abstract}
This paper assesses the implications of a recent discovery (Jenkins 2009) that atomic oxygen is being depleted from diffuse interstellar gas at a rate that cannot be accounted for by its presence in silicate and metallic oxide particles. To place this discovery in context, the uptake of elemental O into dust is considered over a wide range of environments, from the tenuous intercloud gas and diffuse clouds sampled by the depletion observations to dense clouds where ice mantles and gaseous CO become important reservoirs of O. The distribution of O in these contrasting regions is quantified in terms of a common parameter, the mean number density of hydrogen $\nhmean$. At the interface between diffuse and dense phases (just before the onset of ice mantle growth) as much as $\sim 160$~ppm of the O~abundance is unaccounted for. If this reservoir of depleted oxygen persists to higher densities it has implications for the oxygen budget in molecular clouds, where a shortfall of the same order is observed. Of various potential carriers, the most plausible appears to be a form of O-bearing carbonaceous matter similar to the organics found in cometary particles returned by the Stardust mission. The `organic refractory' model for interstellar dust is re-examined in the light of these findings, and it is concluded that further observations and laboratory work are needed to determine whether this class of material is present in quantities sufficient to account for a significant fraction of the unidentified depleted oxygen.

\end{abstract}

\keywords{ISM: abundances --- ISM: molecules --- dust, extinction}

\clearpage

\section{Introduction}
\label{intro}
The depletion of a chemical element in the interstellar gas refers to its observed under-abundance with respect to some standard reference value, resulting from its presumed presence in solid particles. The depletions of the various condensible elements thus yield qualitative evidence for their inclusion in dust, and they also have potential to yield important quantitative information on the probable distribution of those elements in the solid material. It is generally acknowledged that the five-element set \{C, O, Mg, Si, Fe\} contributes most of the mass of the dust in the interstellar medium (ISM), consistent with the finding that models based on silicates and graphitic or amorphous carbon reproduce many of the observed properties of the dust (see Zubko, Dwek, \& Arendt 2004 and Draine \& Li 2007 for recent examples). The depletions are sensitive to interstellar environment and thus provide important clues on the nature of the growth and destruction mechanisms that operate on dust in the ISM.

In the past, the potential of depletion studies to elucidate the nature and evolution of interstellar matter has not been fully realized because of uncertainties and inconsistencies in the measurements, and controversy regarding the best choice of reference abundances. A comprehensive recent study by Jenkins (2009) presents both a detailed review of these issues and an extensive re-analysis of archival data that alleviates them significantly. Data for 17~elements were selected from the literature with careful quality control, and homogenized onto a consistent system of oscillator strengths for the various transitions used to derive column densities. Correlations based on the data were found that enable the rates at which individual elements are depleted with increasing density to be compared. This approach, with its emphasis on {\it rates\/} of depletion rather than absolute values, leads to results that do not depend on the choice of reference abundances. A major conclusion of Jenkins' work is that oxygen is being depleted at a rate that far exceeds that at which it can be incorporated into silicates and metallic oxides. This finding creates a crisis in our understanding of interstellar dust. 

The vast majority of the spectral lines used to measure interstellar depletions lie in the ultraviolet (UV) region of the spectrum. Only relatively low-density interstellar environments can therefore be sampled because the very high UV extinctions arising in dense molecular clouds preclude observation of the background stars needed to provide continuum at these wavelengths. Lines of sight with the highest observed depletions, such as that toward $\zeta$~Oph (the prototypical `high-depletion star'), intercept predominantly atomic gas characterized by observed values for the fraction of hydrogen in molecular form $f({\rm H}_2)\la 0.5$. The uptake of gaseous elements by dust in dense clouds where $f({\rm H}_2)\rightarrow 1$ is observed in a different way, via the occurrence of infrared spectral features identified with solid-state vibrational modes of molecules such as \water, CO and \cod\ (e.g., Gibb \etal\ 2004). These features signal the presence of icy mantles that contain a mix of species formed in situ by surface reactions on dust, together with products of gas-phase chemistry that subsequently condensed. Because of the disparity in sampling technique and a lack of overlap between the respective samples (ice features are not detected in stars studied for element depletions; stars with ice features are too faint in the UV to allow depletions to be evaluated), little attempt has been made to date to integrate the results into a general model for gas-grain interactions in the ISM. 

As well as contributing to our knowledge of dust properties, an integrated approach may be needed to answer a long-standing question concerning the distribution of elemental oxygen in molecular clouds. Observed values and limits on the abundances of important O-bearing species in the gas, such as CO, \water\ and O$_2$, appear insufficient to account for the oxygen not tied up in silicate dust and ices in regions dense enough for the atomic O to be converted into molecules on timescales short compared with cloud lifetimes (e.g., Bergin \etal\ 2000; Goldsmith \etal\ 2000; Larsson \etal\ 2007; Whittet \etal\ 2007; Liseau \& Justtanont 2009): the existence of a missing reservoir of oxygen is implied. The results of Jenkins (2009) suggest the possibility of a causal link between this missing reservoir and depleted oxygen in the diffuse ISM. 

The aim of this paper is to assess the implications of the depletion study of Jenkins (2009) for models of the composition and growth of interstellar dust and the question of repositories for elemental O over the entire range of densities for which relevant observations exist. The following sections examine the depletion of O in diffuse clouds (\S~2) and the transition to dense clouds in which ices and gas-phase CO become significant carriers of elemental O (\S~3). Results are discussed in \S~4 and conclusions are presented in the final section.

\section{The diffuse ISM}
\label{diffuse}
\subsection{Parameters}
The depletion of element X is defined (e.g., Whittet 2003) as
\begin{equation}
D({\rm X}) = \log\,\{{\rm N(X)/N(H)}\} - \log\,\{{\rm {X/H}}\}_{\rm ISM}
\end{equation}
where $N$(X) is the observed gas-phase column density of X, $N({\rm H}) = N({\rm H\,I}) + 2 N({\rm H_2})$ is the total hydrogen column density in the same line of sight, and the term on the right represents the adopted reference abundance of X in the ISM. The fractional abundance of X that is depleted from the gas (and presumed to be in the dust) is then
\begin{equation}
N_{\rm d}({\rm X})/N({\rm H}) = (1 - 10^{D({\rm X})})({\rm X/H})_{\rm ISM}.
\end{equation}
The most appropriate source of reference abundances to use in equations~(1) and (2) is a topic of much discussion and uncertainty (e.g., Whittet 2003; Cartledge \etal\ 2006; Jenkins 2009). The goal is to adopt values that are both accurate and appropriate as a standard for the current ISM. Unevolved early B-type stars provide a logical standard, based on their extreme youth, their general lack of diffusion and convective mixing, and their relatively simple spectra (Przybilla \etal\ 2008). However, solar abundances are often preferred, a choice based implicitly on two questionable assumptions: that the Sun is a normal representative of its stellar type (see Mel\'endez \etal\ 2009 for recent results and caveats), and that the galactic astration cycle has not altered the average interstellar abundances significantly in the 4.6~billion years that have elapsed since the formation of the Sun. In previous work, the accepted value of the oxygen abundance was typically 0.2~dex higher in the Sun compared with B~stars (see Savage \& Sembach 1996 for a review), but recent research has led to a downward revision of the solar value such that results for the two samples are now almost indistinguishable: $\log\,\{{\rm O/H}\}+12=8.69 \pm 0.05$ (solar; Asplund \etal\ 2009) compared with $8.76 \pm 0.03$ (B~stars; Przybilla \etal\ 2008). In the present work I adopt B-type stellar abundances for O and other relevant elements from Table~2 of Przybilla \etal\ (2008)\footnote{Note that Jenkins (2009) adopted solar abundances from Lodders (2003). Coincidentally, the oxygen abundance recommended by Lodders is numerically identical to that adopted here from Przybilla \etal\ (2008); small differences exist in the abundances of Mg, Si and Fe between the two sets, but they have no substantive impact on the results.}.

Jenkins (2009) uses a depletion factor ($\df$) to represent the general degree of depletion in a specific line of sight, such that a larger value of $\df$ implies a stronger depletion for all elements\footnote{See \S~3 of Jenkins (2009) for a detailed explanation of how $\df$ is defined and evaluated.}. $\df$ is chosen to be close to zero in typical lines of sight displaying low depletions, and close to unity in the `prototype' high-depletion region (the $-15$~km/s cloud toward $\zeta$~Ophiuchi). The total range of $\df$ over the entire sample of 243 sight-lines analyzed by Jenkins is $-0.43$ to 1.24. The depletion of element X in a given line of sight is related to the depletion factor by the linear equation
\begin{equation}
D({\rm X}) = B\sx + A\sx(\df\ - z\sx)
\end{equation}
where the slope $A\sx$ represents the propensity of the depletion of X to change with environment as $\df$ varies. Mean values of the coefficients $A\sx$, $B\sx$ and $z\sx$ used in the present work are taken from Table~4 of Jenkins (2009).

\subsection{Silicates and oxides}
The reservoir of oxygen tied up in silicates and oxides is readily quantified. Of the mineral-forming elements, only Mg, Si and Fe need be considered as none of the others are abundant enough to have a significant impact on the depletion of O. Figure~1 plots the solid-phase abundance trends $N_{\rm d}$(X)/$N$(H) against $\df$, combining equations~(2) and (3) with values of the coefficients for each element (X = Mg, Si, Fe) from Jenkins (2009). It should be noted that as the depletions of these elements are generally large and the observations numerous, the fractional errors in the coefficients are generally small, typically $<5\%$. As expected, the solid-phase abundance of each element declines as $\df$ becomes small, and approaches its reference value as $\df$ becomes large. 

The strengths of the mid-infrared Si--O absorption features observed toward reddened stars imply that 80--100\% of all the available Si must be tied up in silicate dust (Mathis 1998). Mg is similar to Si in terms of both abundance and depletion behavior (\fig 1), so it is logical to assume that magnesium silicates MgSiO$_3$ (pyroxene) and Mg$_2$SiO$_4$ (olivine) are prime repositories for both of these elements. Iron silicates are also likely to be present; however, \fig 1 illustrates a known tendency (Fitzpatrick 1996) for Fe to persist in solid form when most of the Mg and Si has been returned to the gas phase in unshielded regions (low values of $\df$), a difference that suggests a substantial reservoir for elemental Fe in a solid form more refractory than silicates. Pure metallic Fe particles are unlikely to exist in the ISM, but oxides such as FeO, Fe$_2$O$_3$ and Fe$_3$O$_4$ are obvious candidates (Jones 1990; Savage \& Sembach 1996; Draine \& Lazarian 1999). A further possibility is that a major fraction of the depleted Fe combines with C instead of O: Weingartner \& Draine (1999) estimate that as much as 60\% of the available Fe might reside in a reservoir attached to small carbonaceous grains.

The solid curves in \fig 1 illustrate a reasonable range for the uptake of oxygen into silicates (and other oxides) as a function of $\df$. The area between the thick curves is constrained by the assumption that all of the depleted metals are oxidized: its upper and lower bounds are set by oxidation ratios O/M = 1.5 and 1.2, respectively, where \hbox{M = Mg+Si+Fe}. Thus, a value of \hbox{O/M = 1.5} is appropriate to MgSiO$_3$ and Fe$_2$O$_3$, the highest oxidation states available\footnote{Silicates will contain additional O if they are hydrated, a possibility discussed in \S~4.2.}, and \hbox{O/M = 1.2} is a reasonable average for a mixture of monoxides (FeO, MgO) and silicates primarily in olivine form (see Kemper \etal\ 2004). The area labeled ``less Fe oxidation" in \fig 1 assumes that the silicate/oxide dust includes all of the depleted Mg and Si but only 40\% of the depleted Fe (with the non-oxidized Fe presumed to be in carbonaceous particles; Weingartner \& Draine 1999); its lower bound (thin solid line) is set by assuming \hbox{O/M = 1.2}, as above. The choice of O/M affects only the vertical scaling of the curves, not their shape. The general pattern of depletion in \fig 1 is consistent with the expectation that grains will be destroyed most easily in low-density regions where $\df \approx 0$. At the opposite end of the scale, the abundance of O incorporated into silicate/oxide dust reaches a plateau value in the range 90--140~ppm at high values of $\df$.

Finally, \fig 1 compares the uptake of oxygen into silicate/oxide dust with the total oxygen depletion. The dashed curves represent upper and lower limits of the trend for O, based on uncertainties in the coefficients (Jenkins 2009) used in equation~(3). This plot further illustrates the conclusion (Jenkins 2009) that the rate at which the oxygen depletion increases with $\df$ far exceeds what can be accommodated in silicates/oxides. Only at the lowest depletions ($\df<0.5$) can silicates and oxides account for most or all of the depleted oxygen in the diffuse ISM. As $\df$ increases toward the high end of its observable range, the total O-depletion continues to rise as the silicate/oxide reservoir levels off; at $\df\approx 1$ as much as half of the depleted O appears to be in some alternative form that I refer to as `unidentified depleted oxygen'.

\section{The dense ISM}
\label{dense}
\subsection{Gaseous CO and ices}
To quantify the uptake of oxygen into solids in denser regions of the ISM, I will concentrate on observations of a single region, the Taurus complex of dark clouds, a well-known site of low-mass star formation. This choice is based on the availability of data on column densities of relevant species, including gaseous CO (Frerking \etal\ 1982; Goldsmith \etal\ 2008) as well as ices (Whittet \etal\ 1988, 2007; Chiar \etal\ 1995) in lines of sight toward a reasonably large sample of confirmed background stars that have well-determined reddening and extinction data (Shenoy \etal\ 2008). Taurus is not necessarily representative of the dense ISM in general, but it currently provides the best opportunity to address the question in hand. 

As the density of a clump of interstellar gas and dust increases, the abundances of gas phase molecules and the adsorption rates of both atoms and molecules from the gas onto the dust increase rapidly, resulting in the efficient conversion of atomic O to gaseous CO and ices (e.g., Hollenbach \etal\ 2009). The observed column densities of both CO gas and major species in the ice (predominantly \water, CO and \cod) correlate approximately linearly with total visual extinction ($\av$) according to the general relation
\begin{equation}
N({\rm X}) = q(\av - \av^{~0})
\end{equation}
(Whittet \etal\ 1988, 2001, 2007), where $q$ is the slope of the correlation, and $\av^{~0}$ is the `threshold' extinction arising in foreground dust in the diffuse ISM and/or in the diffuse outer envelope of the cloud; $N({\rm X})$ is zero to within observational error for $\av < \av^{~0}$. Let us assume that the standard relation $N({\rm H})=k\av$ (where $k \approx 1.9\times 10^{21}~{\rm cm^{-2}\,mag^{-1}}$) is applicable to the dense ISM (see Whittet \etal\ 2007 for discussion of this point). Equation~(4) may then be expressed as an abundance in terms of $N({\rm X})/N({\rm H})$. The total uptake of O into ice mantles on the dust is then given by the sum 
\begin{equation}
N({\rm O})/N({\rm H}) = {1\over k} \sum_i n_i q_i \{1-N_i^{\,0}({\rm H})/N({\rm H})\}
\end{equation}
carried out over all relevant species $i$, where $n_i$ is the number of O atoms in the molecule (i.e., $n_i=1$ for \water\ and CO, and $n_i=2$ for \cod), and $N_i^{\,0}({\rm H})$ is the hydrogen column density corresponding to the observed value of $\av^{~0}$ for that molecule.

\water, CO and \cod\ are by far the most abundance O-bearing ices available to observation in Taurus and other dark clouds (Whittet \etal\ 2007, 2009) and the summation in equation~(5) is therefore limited to these three species in the present work\footnote{The adopted values of $q$ and $\av^{~0}$ for the ices are as listed in \S~4 of Whittet \etal\ (2007).}. Other potential candidates include CH$_3$OH, H$_2$CO, HCOOH and OCS, all of which have been detected in the envelopes of YSOs, but available observations suggest their abundances to be at most a few percent relative to \water\ in quiescent regions of molecular clouds (Chiar \etal\ 1996; Gibb \etal\ 2004). Another potential candidate is O$_2$, but unfortunately its abundance cannot be readily quantified as it lacks sensitive infrared signatures: weak upper limits toward two YSOs (Vandenbussche \etal\ 1999) suggest that its contribution is no more than $\sim 75\%$ of the \cod\ contribution. Laboratory analog experiments and kinetic models developed by Acharyya \etal\ (2007) similarly suggest that O$_2$ is unlikely to be a major constituent of the ices, on the basis its extreme volatility (comparable to that of CO). The presence of O$_2$ at the limiting value suggested by Vandenbussche \etal\ (1999) would increase the uptake of elemental O into ices by $\sim 15\%$. 

Plots of $N({\rm O})/N({\rm H})$ vs.\ $N({\rm H})$ are shown in \fig 2 for ices and gas-phase CO. The data and fits are based on previously published correlations between column densities of the relevant species and $\av$. Each point represents the line of sight to a background field star. The open circles (fit by the dashed curve) denote gaseous CO, and are based on observations of C$^{18}$O, C$^{17}$O and $^{13}$C$^{18}$O by Frerking \etal\ (1982), as discussed by Whittet \etal\ (2007; see their \fig 9)\footnote{Gaseous CO column densities from Frerking \etal\ (1982) are in excellent agreement with values extracted from the extensive recent CO survey of the Taurus cloud by Goldsmith \etal\ (2008); these results will be presented in a future paper.}. The filled circles (fit by the solid curve) denote the summation of oxygen from gaseous CO and O-bearing ices. Ice column densities are taken from Table~2 of Whittet \etal\ (2007), omitting stars that lack data for \water\ or \cod\ (10 out of 13 are plotted). In a few lines of sight where no direct measurement of gaseous CO is unavailable from Frerking \etal, a notional value is calculated from the $N$(H) vs.\ $N$(CO) correlation for other stars.

The form of \fig 2 may be understood as follows. It is clear from the locus and properties of the Taurus complex (Shenoy \etal\ 2008) that almost all of the interstellar material detected toward background field stars arises in the complex itself, and the column densities observed toward a given star are therefore independent of its distance (provided only that it is behind the cloud). The column density $N$(H) is therefore a proxy for the particle number density. Gaseous CO and ices becomes detectable as $N$(H) increases above the threshold value for each species in turn, owing to increased collision rates and attenuation of the external radiation field within the cloud; their abundances increase steadily and then level off as $N$(H) becomes large. Extrapolation of these trends suggests that, deep within the cloud, gaseous CO and ices account for $\sim 43$~ppm and $\sim 116$~ppm of elemental O, respectively, values that leave $\ga 50\%$ of the oxygen not in silicate/oxide dust unaccounted for if the assumed reference abundance is appropriate. 

\subsection{The Diffuse-Dense ISM Interface}
In general, column densities toward stars studied for depletion (\S~2) represent the accumulation of material over a long path length that may include several discrete concentrations, whereas those toward the dark cloud discussed above arise predominantly in the cloud itself. The goal of this section is to attempt to describe both sets of data in terms of a common parameter, chosen to be the mean hydrogen number density, $\nhmean = N({\rm H})/L$, where $L$ is the length of the column. The real physical density of the ISM at points along the column to a distant star will typically be subject to enormous variation, but Spitzer (1985) showed, nevertheless, that $\nhmean$ provides meaningful insight into physical conditions. Lines of sight with low values of $\nhmean$ ($< 0.2$~cm$^{-3}$) are typically dominated by warm low-density gas, those with intermediate values ($\sim 0.7$~cm$^{-3}$) by `standard' diffuse H\,I clouds, and those with higher values ($> 3$~cm$^{-3}$) by denser diffuse clouds or complexes (corresponding to the `diffuse molecular' and `translucent' classifications proposed by Snow \& McCall 2006). Values of $\nhmean$ for stars studied by Jenkins (2009) occupy the range \hbox{0.02\,--\,5~cm$^{-3}$} (ignoring a few outliers): his sample is thus dominated by low and intermediate density regimes.

It is informative to compare the line of sight to the high-depletion star $\zeta$~Oph, from the Jenkins study with that to the star with the lowest extinction amongst those with ice detections (HD~29647; Whittet \etal\ 1988). Relevant data are listed in Table~1. A comparison of $N$(H) and $\nhmean$ data with the model clouds of Spitzer (1985) and Snow \& McCall (2006) suggests that $\zeta$~Oph samples a diffuse molecular cloud (as expected), and HD~29647 samples material near the interface of translucent and dense molecular clouds. In the latter case, this result is consistent with the locus of HD~29647 behind the outer envelope of \hbox{TMC--1} (Crutcher 1985), with the observed weakness of its detected ice band, and with the fact that its $\av$ value is similar to the \water-ice threshold extinction $\av^{~0}\approx 3.3$~mag in Taurus (Whittet \etal\ 2001). An increase in $\nhmean$ from $\sim 4$~cm$^{-3}$ ($\zeta$~Oph) to $\sim 13$~cm$^{-3}$ (HD~29647) may thus be taken to represent the transition from a high-depletion region of the diffuse ISM (where gaseous CO is detected but accounts for only \hbox{$\sim 1$\,--\,2~ppm} of the available O) to the diffuse/dense ISM interface where the abundances of both gaseous CO and ices are increasing rapidly with density. 

A tight linear correlation is demonstrated between $\nhmean$ and $\df$, given by
\begin{equation}
\df = 0.772 + 0.461\log\,\nhmean
\end{equation}
 (Jenkins 2009), and this equation enables a simple conversion of the depletion data: \fig 3 includes a representation of the mean trends for silicates/oxides and elemental O from \fig 1, replacing $\df$ with $\nhmean$ as the abscissa via equation~(6). To replot the trends in \fig 2 in the same way requires a value to be adopted for the effective column length $L$. Factoring the individual distances to the background stars used to establish the correlations into this conversion would introduce random scatter arising from the arbitrary length of the segment of each column lying behind the cloud (where no significant contribution to the column density is expected). Instead I adopt a single value for $L$, chosen to be the distance to HD~29647 (Table~1). This star is situated close to the rear boundary of the Taurus complex (Whittet \etal\ 2004) and appears to be an appropriate delineator between diffuse and dense regimes for reasons discussed above. Adopting a different value for $L$ would simply introduce a horizontal displacement of the molecular zones (ices and CO) in \fig 3 relative to the depletion curves.

\section{Discussion}
\label{discussion}
\subsection{Overview}
The purpose of Figure~3 is to present a schematic overview of the distribution of oxygen between major reservoirs over a wide range of interstellar environments, combining the trends determined from the observations of diffuse and dense regions of the ISM described in the previous sections. The total range of the ordinate is set to the adopted reference abundance for interstellar oxygen (575~ppm; Przybilla \etal\ 2008). The vertical dashed line denotes the effective observational limit on depletion studies imposed by the high extinction of UV flux from stars that sample dense regions. Extrapolation of the trend for silicates and oxides from the diffuse ISM to higher density is justified, based on the non-volatile nature of these materials and spectroscopic detection of silicates in dense clouds (e.g., Bowey \etal\ 1998). A much more speculative extrapolation of the unidentified depleted oxygen (UDO) region is also shown. 

The well-documented problem of the oxygen budget in the dense ISM is evident in \fig 3: the combined contributions of gaseous CO, ices and silicate/oxide dust account for no more than $\sim 300$~ppm of the O at high density. So even if a significantly lower reference abundance is adopted for O (e.g., the solar value of 490~ppm proposed by Asplund \etal\ 2009) there remains a substantial shortfall. As previously noted, it does not appear possible to account for this discrepancy in terms of known interstellar gas-phase molecules (\S~1) or additional ice constituents (\S~3). This finding is inconsistent with the results of time-dependent models developed by Hollenbach \etal\ (2009), which predict that essentially all of the O not in silicates/oxides should eventually be processed into gaseous CO and ices in shielded regions deep within a molecular cloud. The Taurus cloud may still be evolving toward such a state, but it nevertheless seems unlikely that these two reservoirs can entirely solve the problem. The diffuse-ISM data imply that an additional reservoir of depleted oxygen was already present before the ice formation process began, and this could have limited the availability of atomic O to make ices. If the reservoir represented by UDO persists in the dense ISM (\fig 3), it contains enough O to explain the shortfall.

Only two elements, H and C, are abundant enough to combine with O in quantities sufficient to account for the depleted O not in silicates and oxides in the diffuse ISM (assuming O$_2$ is not a major player for reasons already discussed). The remainder of this section attempts to address two key questions: What constraints can be placed on the nature of the unidentified depleted oxygen in the diffuse ISM, and how might this reservoir evolve as matter is cycled between diffuse and dense phases of the ISM?

\subsection{Water-ice and hydrated silicates}
Solid \water\ is an obvious H-bearing candidate, but the intrinsic strength of the absorption feature produced by its \hbox{O--H} vibrational mode at 3.0\mic\ provides a rather stringent test for its presence as a component of the dust in the diffuse ISM. Whittet \etal\ (1997) set a limit of \hbox{$<2$~ppm} for the abundance of O in this form toward the highly reddened star Cyg~OB2 no.~12. This result is consistent with the `threshold effect' observed in the dense ISM (\S~3.1), which implies that ices are lacking in the diffuse outer layers of molecular clouds. Jenkins (2009) notes that thick ice mantles on large ($>1$\mic) grains would be hard to detect spectroscopically, in which case this limit might be relaxed somewhat; selective desorption of mantles from smaller grains might then be understood in terms of transient heating induced by absorption of individual energetic photons. However, the grain size distribution is regulated by shattering caused by grain-grain collisions in shock waves (Jones \etal\ 1996; Slavin \etal\ 2004), which results in a rapid (power-law) decline in numbers with increasing radius (see \fig 19 of Zubko \etal\ 2004 for examples of size distribution functions used in models that successfully match the observed extinction and emission of the dust). Moreover, even thick ice mantles on micron-sized grains seem unlikely to survive the rigours of photodesorption (Westley \etal\ 1995) and sputtering (Jones \etal\ 1996) in unshielded regions of the ISM on time scales long enough to constitute a significant reservoir of O in such environments. One further scenario to consider is the presence of a substantial interstellar population of macroscopic icy bodies (free-flying comets or planetesimals); but unless they were composed of almost pure \water\ it would not be possible to solve the oxygen problem without exceeding the availability of other elements such as C, N, Mg, Si and Fe that such bodies would presumably contain. 

The uptake of O into silicates (\fig 1) will be enhanced if interstellar silicates are appreciably hydrated. This possibility may be explored by observations of \hbox{O--H} features that arise in hydrated silicates in the wavelength range 2.6--2.8\mic\ (unfortunately, a spectral region unavailable to ground-based observation). Whittet \etal\ (1997) reported a tentative detection of a weak hydrated silicate feature in a spectrum of Cyg~OB2 no.~12 obtained with the Infrared Space Observatory, but this was later found to be an artefact (Whittet \etal\ 2001). The corrected spectrum sets a limit of \hbox{$<5$~ppm} for the abundance of OH groups in silicate dust toward this star. If this line of sight is representative, hydration has negligible impact on the depletion of O.

\subsection{Carbon}
The possibility remains that depleted oxygen is somehow sequestered into carbonaceous dust. This might occur as the result of simple adsorption reactions in the diffuse ISM, in which case very small ($<0.01$\mic) grains and macromolecules should be the primary repositories as they contain most of the available surface area (e.g., Weingartner \& Draine 1999). The leading candidates are polycyclic aromatic hydrocarbons (PAHs), responsible for the well-known aromatic spectral emission features in the infrared, and small graphitic particles, presumed to account for the 218~nm ultraviolet absorption feature; these populations are estimated to contain $\sim 20$~ppm and $\sim 50$~ppm of the available carbon, respectively, with a further $\sim 100$~ppm residing in larger grains (Mathis 1996; Tielens 2008). The attachment might be purely physical or it might involve chemical bonding with surface atoms. The valence structure of O prevents it from substituting for a C atom within an aromatic ring of a PAH molecule or graphitic layer, and chemical attachment is thus limited to defect sites or sites at the outer edges of peripheral rings (Hudgins \etal\ 2005; Ajayan \& Yakobson 2006).\footnote{A further caveat is provided by the laboratory work of Betts \etal\ (2006) and Eichelberger \etal\ (2007), which suggests that the most likely outcome of an encounter between an O atom and an aromatic or aliphatic hydrocarbon is abstraction of a C atom to form gaseous CO; but as the experiments were performed at room temperature, their relevance to the ISM may be questionable.} Surface adsorption at a level sufficient to explain the excess depletion of O at $\df\sim 1$ \hbox{($\sim 80$\,--\,160~ppm;} \fig 1) would require O/C $> 1$ in the very small grains, which is clearly unrealistic. Sufficient depletion seems possible only if the O is incorporated within the structure of the grain material, and, specifically, within the larger particles that contain most of the mass and volume of the dust; in this case, O/C ratios as low as $\sim 0.5$ (averaged over all grain sizes) could suffice. 

Carbonaceous interstellar dust originates in the outflows of evolved stars (`stardust') and in the ISM itself as matter is cycled through molecular clouds and sites of star formation. The composition of stardust injected from a given source is determined by the O/C ratio of the stellar envelope in which it condenses: O-rich and C-rich dust has been observed to coexist in some outflows (e.g., Perea-Calder\'on \etal\ 2009), but this appears to be relatively rare and the mixing macroscopic (the simultaneous presence of products from distinct O-rich and C-rich phases of dust production) rather than microscopic (O embedded in individual C-rich dust grains). There seems to be little scope for substantial O-content in carbonaceous stardust prior to processing in the ISM.

Molecular clouds appear far more promising as environments in which mixing of O into C-rich dust might occur. Grain-grain coagulation and ice-mantle growth proceed in tandem, and the smaller carbonaceous grains (PAHs, graphite) are expected to become embedded in the mantles of larger grains: ultimately, the surfaces of essentially all pre-existing C-rich dust (large and small) will come into intimate contact with the ices in the densest regions. The ices themselves contain newly-manufactured C-bearing species such as CO, \cod, CH$_3$OH and CH$_4$ that have the potential to be converted into more robust carbonaceous grain material when stars switch on within the clouds. As the net composition of the ice is strongly O-rich, the products of energetic processing are expected to be oxygenated. Laboratory experiments demonstrate, for example, that PAHs embedded in interstellar ice analogs acquire stable O-bearing side groups when subject to proton bombardment or UV photon irradiation (Bernstein \etal\ 1999, 2002b, 2003; Ashbourn \etal\ 2007). More generally, it is well known that irradiation of ices in the laboratory under simulated interstellar conditions leads to the production of a variety of complex organic molecules and refractory residues with significant oxygen content (e.g., Jenniskens \etal\ 1993; Bernstein \etal\ 1995, 2002a; Greenberg \etal\ 1995; Pendleton \& Allamadola 2002; Mu\~noz Caro \& Schutte 2003). These products are likely to contribute to the organic inventories of primitive bodies such as comets and asteroids in the Solar System (e.g., Pendleton 1997; Kerridge 1999; Ashbourn \etal\ 2007), and it is notable that samples of cometary dust returned by the Stardust mission and interplanetary dust collected in the stratosphere contain refractory organics with O/C ratios as high as $\sim 0.5$ (Sandford \etal\ 2006). 

The hypothesis that star formation drives widespread production of carbonaceous dust with O/C $\sim 0.5$ seems plausible and in accord with general models for the evolution of dust in the Galactic ecosystem (e.g., Zhukovska \etal\ 2008 and references therein). Cycling of this material from the production sites into lower-density regions could effectively account for the `missing' reservoir of depleted oxygen. The composition of an initially O-rich residue will change on a timescale $\ga 3 \times 10^7$~years (Jenniskens \etal\ 1993) in response to long-term irradiation, until ultimately reduced to amorphous carbon following a gradual release of O and other heteroatoms from the structure. The variation in O-depletion with density in the diffuse ISM (\fig 3) may then be understood simply in terms of different degrees of desorption in different environments. But if some grains are cycled back into molecular clouds before the desorption process is complete, the organics that are retained will become a stable second-generation reservoir of O-bearing dust in the dense ISM.

This scenario is also consistent with considerations regarding the longevity of carbonaceous dust in the diffuse ISM. The large grains responsible for extinction at visible wavelengths must include carbonaceous dust, in order to satisfy constraints on the opacity per unit dust mass; amorphous or partially graphitized carbon are obvious and logical candidates (e.g., Mathis 1996; Zubko \etal\ 2004). An interstellar source of carbonaceous dust is implied by estimates of its production timescale in the outflows of C-rich evolved stars, which far exceed estimates of its destruction timescale in interstellar shocks (Jones \etal\ 1996). Moreover, recent work shows that amorphous carbon is less stable to destruction by shocks than previously thought (Serra D\'iaz-Cano \& Jones 2008), further emphasizing the need for an effective production mechanism. The laboratory simulations and calculations of Jenniskens \etal\ (1993) suggest that the cycling of interstellar solids from ices to organic residues to amorphous carbon is indeed an efficient process capable of providing much of the carbonaceous matter in the diffuse ISM as a natural by-product of star formation. 

\subsection{An organic refractory component of interstellar dust?}
The new result on interstellar depletions discussed in this paper motivate a reassessment of the proposal (e.g., Li \& Greenberg 1997) that organic refractory matter is a significant component of interstellar dust. The Li \& Greenberg model successfully explains many of the observed properties of interstellar dust in terms of separate populations of silicate grains bearing organic refractory mantles, small graphite particles and PAHs, all essentially within the limits set by abundance constraints. However, the feasibility of the model is severely challenged by spectroscopic observations in the infrared. The search for spectroscopic evidence of organic refractory matter in the diffuse ISM has focused in much previous work on the identity of an interstellar absorption feature at 3.4\mic\ (the aliphatic \hbox{C--H} stretching mode) which is well matched by laboratory spectra of appropriate analogs (Greenberg \etal\ 1995; Pendleton \& Allamandola 2002). However, this feature is surprisingly weak or absent in denser environments where the organics are thought to be forming (Pendleton \& Allamandola 2002; Shenoy \etal\ 2003). Searches for a spectropolarimetric counterpart to the 3.4\mic\ absorption, expected if the carrier is a mantle on aligned silicate cores, have set significant upper limits (Adamson \etal\ 1999; Chiar \etal\ 2006) that suggest a more likely origin for the observed feature in the small-grain population (which generally fails to align). Finally, searches for features expected to occur in organic refractory matter at other wavelengths (including, crucially, those arising from oxygen bonding such as \hbox{O--H} and C=O) have also yielded negative results (Schutte \etal\ 1998; Pendleton \& Allamadola 2002). The latter authors conclude that organic refractory material in the diffuse ISM is composed predominantly of aromatic and aliphatic hydrocarbons with minimal oxygen content (O/C $\sim 0.015$).

Do these objections rule out organic refractory matter as a significant carrier of unidentified depleted oxygen? The available observations may not provide a conclusive test. There are several issues to consider, including sample selection and size, blending of features, and the availability of data on band strengths in appropriate analogs. As the observed 3.4\mic\ feature is intrinsically quite weak, very large dust columns are needed for its detection, and to date the sample has been limited to lines of sight toward the Galactic Center ($\av \approx 30$~mag) and a few other luminous stars with $5\la\av\la 20$~mag (Sandford \etal\ 1995; Chiar \etal\ 2000, 2002). Moreover, only two (the Galactic Center and the B-type hypergiant Cyg~OB2 no.\,12) have 3.4\mic\ data of sufficient quality to enable meaningful profile comparisons with laboratory spectra (Pendleton \& Allamadola 2002). The Galactic Center is obviously a very complicated line of sight, subject to galactocentric abundance trends, and displaying evidence for a molecular-cloud component to its absorption superposed on the diffuse-ISM component. Thus, Cyg~OB2 no.\,12 ($\av \approx 10$~mag) is generally adopted as the prototype target for infrared studies of the diffuse ISM (e.g., Whittet \etal\ 1997; Pendleton \& Allamadola 2002), but this is by default rather than by design; it may indeed be highly atypical in comparison with lines of sight used to study depletions, which all have much lower extinctions. Moreover, the observed 3.4\mic\ feature may not in fact arise in a single grain material but may be a blend of absorptions in (e.g.) hydrogenated amorphous carbon as well as organic matter (with only the latter containing significant oxygen). If this is the case, constraints based on the apparent strengths of features at other wavelengths relative to that at 3.4\mic\ become more difficult to interpret. 

The intrinsic and relative strengths of the various features seen in the spectra of laboratory organic residues vary from sample to sample and are highly dependent on initial composition, physical conditions, and the nature and degree of the energetic processing (e.g., Schutte 1988; Jenniskens \etal\ 1993). The \hbox{5\,--\,8\mic} spectral region typically contains a complex blend of features arising in several vibrational modes, most notably a distinctive absorption edge near 5.85\mic\ seen in many samples (e.g., \fig 1 of Greenberg \etal\ 1995). This edge is associated with the onset of carbonyl \hbox{($>$C=O)} absorption in the residues. Interestingly, Jenniskens \etal\ found relatively strong carbonyl absorption in a highly irradiated sample that displays very weak 3.4\mic\ absorption, concluding that the presence of an adjacent carbonyl group suppresses the \hbox{C--H} resonance.

The 5.8\mic\ region of the spectrum is less well studied in astronomical spectra than the 3.4\mic\ region because of higher telluric absorption that precludes ground-based observation. Moreover, interpretation of data for the best studied target (the Galactic Center) is complicated by the presence of overlapping features attributed to the diffuse ISM and molecular clouds (Chiar \etal\ 2000). However, weak 5.85\mic\ absorptions have been detected in the spectra of several highly reddened Wolf-Rayet stars (Schutte \etal\ 1998). The strengths of the observed features are consistent with an oxygen abundance of \hbox{$\sim 3$\,--\,12~ppm} in carbonyl groups along these lines of sight, which represents $<10\%$ of the missing O. It should be noted, however, that this calculation depends on a measure of the band strength for the feature ($2\times 10^{-17}$~cm/molecule; Wexler 1967) that may have little relevance to interstellar materials. It would clearly be important to determine band strengths for this and other features in samples of organic refractory matter produced under simulated interstellar conditions. 

\section{Conclusions}
The uptake of elemental oxygen into interstellar dust has been quantified over environments ranging from the tenuous intercloud gas and diffuse clouds sampled by observations of elemental depletions to dense regions where ices and gaseous CO become important reservoirs of O. At the interface between diffuse and dense phases (just before the onset of ice-mantle growth) as much as $\sim 160$~ppm of the depleted O is unaccounted for (\fig 3), a conclusion insensitive to the choice of reference abundances. If this reservoir of depleted oxygen persists to higher densities it has implications for the oxygen budget in molecular clouds as well. Of various potential carriers, the most plausible appears to be a form of O-bearing carbonaceous matter similar to the organics found in cometary particles returned by the Stardust mission. The results of the present work thus provide presumptive evidence for an organic refractory component of interstellar dust with significant oxygen content.

The `organic refractory' model for interstellar dust (Li \& Greenberg 1997) is reviewed in the light of these findings. Infrared spectroscopy places constraints on the abundances of O-bearing functional groups in the organics, but further observations and laboratory work are needed to determine whether this class of material is present in quantities sufficient to account for an appreciable fraction of the missing oxygen. It will be important in the future to obtain spectra of the highest quality for stars in lines of sight that sample diffuse-ISM dust over a range of visual extinctions, especially in the waveband containing the 5.85\mic\ carbonyl feature as it should provide the most sensitive quantitative test for oxygenated organics. Both the Stratospheric Observatory for Infrared Astronomy and the James Webb Space Telescope will have instruments well matched to this task. Further laboratory work should also be done to measure the band strengths of carbonyl and other vibrational features in organic refractory residues produced over a range of realistic conditions.

\acknowledgments
Financial support for this research was provided by the NASA Exobiology and Evolutionary Biology program (grant NNX07AK38G) and the NASA Astrobiology Institute (grant NNA09DA80A). I am grateful to Max Bernstein, Perry Gerakines and an anonymous referee for helpful comments.

\clearpage


\begin{deluxetable}{lccccccc}
\tablecaption{Summary of data for two benchmark lines of sight.}
\tablewidth{0pt} 
\tablehead{\colhead{Star} & \colhead{Distance} & \colhead{$A_V$} & \colhead{$N$(H)} & 
\colhead{$N$(CO)} & \colhead{$\nhmean$} & \colhead{$f$(H$_2$)} & \colhead{Notes}\\ 
& (pc) & (mag) & (cm$^{-2}$) & (cm$^{-2}$) & (cm$^{-3}$)}
\startdata 
$\zeta$~Oph & 110\,--\,140 & 0.99 & $1.4\times 10^{21}$ & $2.4\times 10^{15}$ & 4.1 & 0.63 & 1  \\
HD~29647    & $170\pm 20$  & 3.63 & $6.9\times 10^{21}$ & $\sim 4\times 10^{17}$~ & 13  & $\sim 1$? & 2 \\ 
\enddata

\tablenotetext{~}{\\ \small (1)~For consistency, the distance of 110~pc listed by Jenkins (2009) was used to estimate $\nhmean$; parallax data suggest a somewhat larger value of $140 \pm 14$~pc (Perryman \etal\ 1997). Reddening and column densities are from Liszt \etal\ (2009) and references therein; $\av=3.1E_{\rm B-V}$ is assumed.} 
\tablenotetext{~}{\\ \small (2)~The adopted photometric distance (Crutcher 1985) is consistent with parallax data (Perryman \etal\ 1997). $\av$ is from Whittet \etal\ (2001), and the standard mean ratio $N(H)/\av= 1.9\times 10^{21}$~cm$^{-2}$~mag$^{-1}$ is used to estimate $N$(H). $N$(CO) is from Crutcher (1985). $f$(H$_2$) is unknown but presumed to be close to unity.}
\end{deluxetable}

\clearpage


\begin{figure}
\centering
\includegraphics[width=12cm, angle=0]{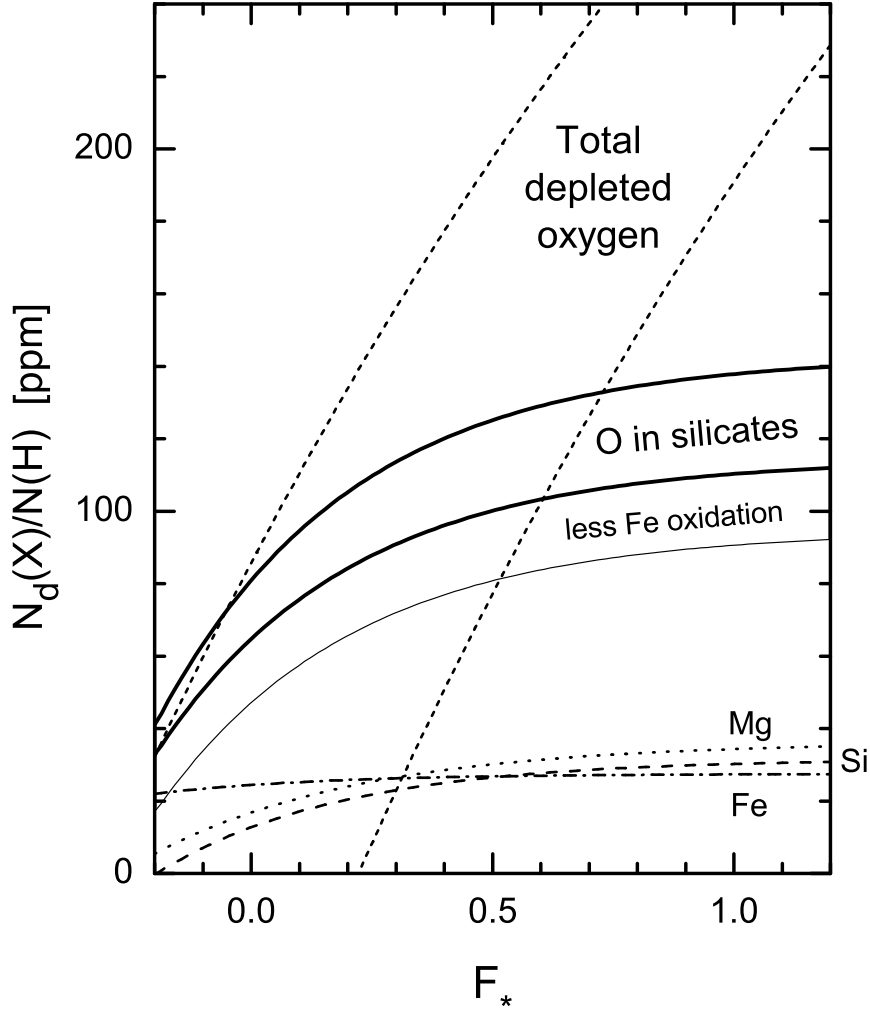}
\caption{\small Plot of solid-phase abundance $N_{\rm d}$(X)/$N$(H) vs.\ depletion factor $\df$ for the principal elements depleted into silicate/oxide dust in the ISM (X = O, Mg, Si, or Fe). Abundances in this and subsequent figures are presented in parts per million (ppm). The dotted, dashed and dot-dashed curves toward the bottom of the plot denote Mg, Si and Fe, respectively, constructed from empirical fits to depletion data from Jenkins (2009). The area labelled `O in silicates' represents the probable range of O depletion into silicates and oxides, assuming oxidation of all depleted Mg, Si and Fe, with O/M = 1.5 and 1.2 (upper and lower bounds, respectively, thick solid curves) where M = Mg+Si+Fe. The thin solid line indicates a possible downward revision of the lower bound, assuming that the silicate/oxide dust includes all of the depleted Mg and Si but only 40\% of the depleted Fe (Weingartner \& Draine 1999). Upper and lower limits on the total depleted oxygen (dashed curves) are based on empirical fits from Jenkins (2009). \label{fig1}}
\end{figure}
\clearpage


\begin{figure}
\centering
\includegraphics[width=14cm, angle=0]{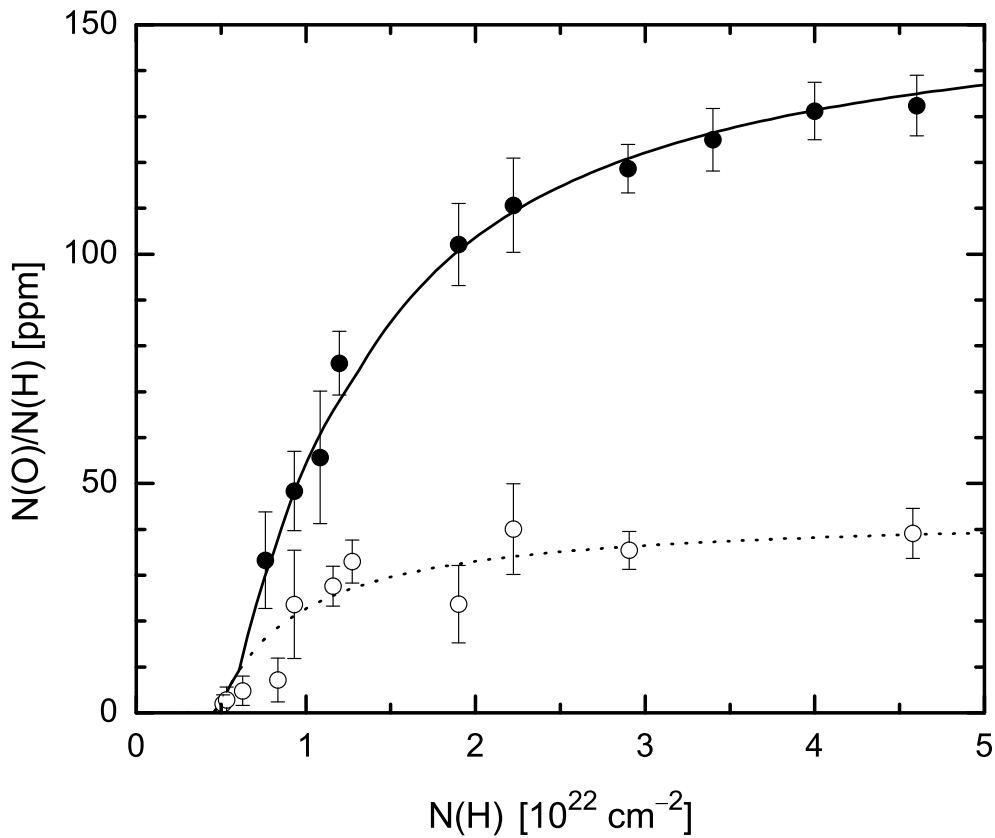}
\caption{\small The uptake of oxygen into gas phase CO and ices with increasing column density in a prototypical dense cloud (Taurus). The data and fits are based on previously published correlations between column densities of the relevant species and visual extinction toward background field stars (Whittet \etal\ 2007 and references therein). The open circles (fit by the dashed curve) denote gas phase CO; the filled circles (fit by the solid curve) denote the summation of gaseous CO and the primary O-bearing species in the ices (\water, CO and \cod). See text (\S~3.1) for further discussion. \label{fig2}}
\end{figure}
\clearpage


\begin{figure}
\centering
\includegraphics[width=13cm, angle=0]{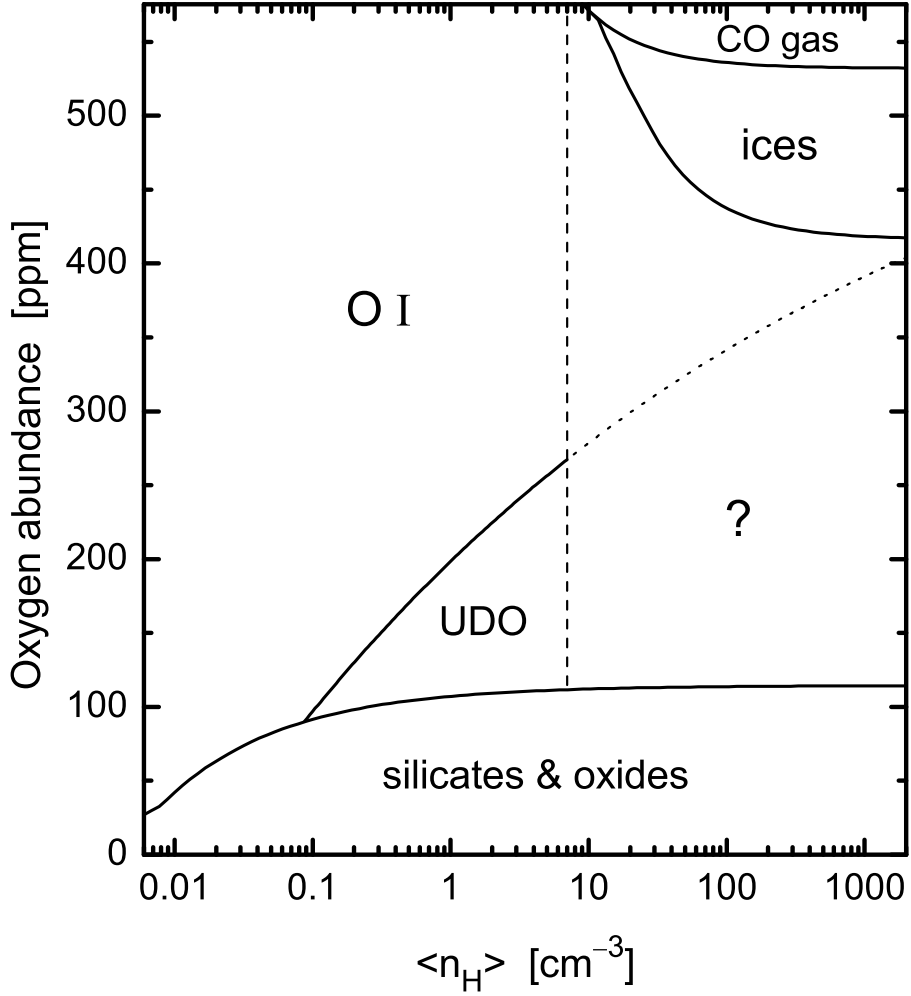}
\caption{\small A schematic overview of the distribution of oxygen between major reservoirs over a wide range of interstellar environments, combining observations of element depletion in the diffuse ISM with those of ices and gaseous CO in the prototypical dense cloud. Ticks on the ordinate correspond to increments of 50~ppm, and the total range is set to the adopted reference abundance (575~ppm). The abscissa represents mean line-of-sight number density (see \S~3.2) and is not the same as the real physical number density at which important processes occur. The vertical dashed line denotes the effective observational limit on depletion studies imposed by the UV opacity of the ISM at higher densities. The ``silicates \& oxides" region is bounded by the median of the shaded and hatched curves in \fig 1. The ``CO gas" and ``ices" regions are based on the fits to data shown in \fig 3 (inverted for display). The area labelled UDO (unidentified depleted oxygen) is bounded by the median of the upper and lower curves for O in \fig 1; a speculative extrapolation to densities higher than those sampled by the relevant observations is also shown (dotted curve). \label{fig3}}
\end{figure}
\clearpage

\end{document}